\documentclass[longauth]{aa}  

\usepackage{graphicx}
\usepackage[varg]{txfonts}
\usepackage{lipsum}
\usepackage{lscape}
\usepackage{placeins}
\usepackage{mathrsfs}
\usepackage{amssymb, amsmath}

\usepackage[
 bookmarks=true,
 pdfnewwindow=true,
 colorlinks=true,
 linkcolor=blue,
 citecolor=blue,
 filecolor=blue,
 urlcolor=blue,
 final=true,
]{hyperref}

\usepackage[
 print-unity-mantissa=false,
 exponent-product=\ensuremath{\cdot},
 uncertainty-mode = separate,
 separate-uncertainty-units = single,
]{siunitx}

\DeclareSIUnit{\year}{yr}
\DeclareSIUnit{\erg}{erg}
\DeclareSIUnit{\deg}{deg}
\DeclareSIUnit{\photons}{ph}
\DeclareSIUnit{\counts}{cts}
\DeclareSIUnit{\photoelectrons}{p.e.}
\DeclareSIUnit{\parsec}{pc}
\DeclareSIUnit{\gauss}{G}
\DeclareSIUnit{\solarmass}{\text{\ensuremath{\textnormal{M}_\odot}}}
\newcommand{\Li}{\mathscr{L}}

\newcommand{\Po}{\mathcal{P}}
\newcommand{\Poi}{\ensuremath{\text{Poi}}}
\def \d {\mathrm{d}}

\begin{document}

\title{CTAO LST$-$1 observations of magnetar SGR~1935$+$2154:\\Deep limits on sub-second bursts and persistent tera-electronvolt emission}

\author{
K.~Abe\inst{1} \and
S.~Abe\inst{2} \and
A.~Abhishek\inst{3} \and
F.~Acero\inst{4,5} \and
A.~Aguasca-Cabot\inst{6} \and
I.~Agudo\inst{7} \and
C.~Alispach\inst{8} \and
D.~Ambrosino\inst{9} \and
F.~Ambrosino\inst{10} \and
L.~A.~Antonelli\inst{10}
% ...
}
\institute{
Department of Physics, Tokai University, 4-1-1, Kita-Kaname, Hiratsuka, Kanagawa 259-1292, Japan
\and Institute for Cosmic Ray Research, University of Tokyo, 5-1-5, Kashiwa-no-ha, Kashiwa, Chiba 277-8582, Japan
\and INFN and Università degli Studi di Siena, Dipartimento di Scienze Fisiche, della Terra e dell'Ambiente (DSFTA), Sezione di Fisica, Via Roma 56, 53100 Siena, Italy
\and Université Paris-Saclay, Université Paris Cité, CEA, CNRS, AIM, F-91191 Gif-sur-Yvette Cedex, France
\and FSLAC IRL 2009, CNRS/IAC, La Laguna, Tenerife, Spain
\and Departament de Física Quàntica i Astrofísica, Institut de Ciències del Cosmos, Universitat de Barcelona, IEEC-UB, Martí i Franquès, 1, 08028, Barcelona, Spain
\and Instituto de Astrofísica de Andalucía-CSIC, Glorieta de la Astronomía s/n, 18008, Granada, Spain
\and Department of Astronomy, University of Geneva, Chemin d'Ecogia 16, CH-1290 Versoix, Switzerland
\and INFN Sezione di Napoli, Via Cintia, ed. G, 80126 Napoli, Italy
\and INAF - Osservatorio Astronomico di Roma, Via di Frascati 33, 00040, Monteporzio Catone, Italy
% ...
}

\titlerunning{Upper limits on sub-second tera-electronvolt burst emission of SGR~1935$+$2154}
\authorrunning{The CTAO-LST Collaboration}

\date{Received May 9, 2025; accepted October 29, 2025}

\abstract
{The Galactic magnetar SGR~1935$+$2154 has exhibited prolific high-energy (HE) bursting activity in recent years.}
{Investigating its potential tera-electronvolt counterpart could provide insights into the underlying mechanisms of magnetar emission and very high-energy (VHE) processes in extreme astrophysical environments. 
We aim to search for a possible tera-electronvolt counterpart to both its persistent and sub-second-scale burst emission.}
{We analysed over $\SI{25}{\hour}$ of observations from the Large-Sized Telescope prototype (LST$-$1) of the Cherenkov Telescope Array Observatory (CTAO) during periods of HE activity from SGR~1935$+$2154 in 2021 and 2022 to search for persistent emission.
For bursting emission, we selected and analysed nine $\SI{0.1}{\second}$ time windows centred around known short X-ray bursts, targeting potential sub-second-scale tera-electronvolt counterparts in a low-photon-statistics regime.}
{While no persistent or bursting emission was detected in our search, we establish upper limits for the tera-electronvolt emission of a short magnetar burst simultaneous to its soft gamma-ray flux.
Specifically, for the brightest burst in our sample, the ratio between tera-electronvolt and X-ray flux is $\lesssim \num{e-3}$.}
{The non-detection of either persistent or bursting tera-electronvolt emission from SGR~1935$+$2154 suggests that if such components exist, they may occur under specific conditions not covered by our observations.
This aligns with theoretical predictions of VHE components in magnetar-powered fast radio bursts and the detection of $\si{\mega\electronvolt}-\si{\giga\electronvolt}$ emission in giant magnetar flares.
These findings underscore the potential of magnetars, fast radio bursts, and other fast transients as promising candidates for future observations in the low-photon-statistics regime with Imaging Atmospheric Cherenkov Telescopes, particularly with the CTAO.}

\keywords{
    \href{http://astrothesaurus.org/uat/992}{Magnetars (992)} --
    \href{http://astrothesaurus.org/uat/1471}{Soft gamma-ray repeaters (1471)} --
    \href{http://astrothesaurus.org/uat/1858}{Methods: Data Analysis (1858)} --
    \href{http://astrothesaurus.org/uat/628}{Gamma-ray astronomy (628)} --
    \href{http://astrothesaurus.org/uat/2109}{Time domain astronomy (2109)} --
    \href{http://astrothesaurus.org/uat/1851}{Transient sources (1851)}
}

\maketitle

\nolinenumbers
\section{Introduction}
\label{sec:Introduction}
Magnetars are a class of neutron star (NS) characterised by extreme variability in the X-ray and soft gamma-ray bands, showing activity that ranges from bursts of a few milliseconds to prolonged outbursts that last for months \citep{kaspi_magnetars_2017}.
Most magnetars are isolated and slowly rotating (period $P \sim\qtyrange{1}{12}{\second}$), and possess a strong magnetic field of $\qtyrange{1E14}{1E15}{\gauss}$, about three orders of magnitude higher than that of regular NSs \citep{olausen_magnetarcatalog_2014}.
The magnetic field acts as the main energy source and the emission can be distinguished in two categories: the persistent emission from the magnetar (and the surrounding nebula if present) and the bursting emission composed of outburst episodes with bursts or flares on different timescales, likely caused by crustal fractures and/or rearrangements of the magnetosphere \citep{beloborodov_corona_2007, mereghetti_magnetars_2015}.
There are about $30$ known magnetars in the Galaxy and Magellanic Clouds \citep{olausen_magnetarcatalog_2014}\footnote{\url{http://www.physics.mcgill.ca/~pulsar/magnetar/main.html}}.
Many of them spend most of the time in a state of low luminosity, and only during outburst episodes does their luminosity rapidly increase \citep{mereghetti_magnetars_2015, kaspi_magnetars_2017, coti_outbursts_2018}.
These episodes can last from a few weeks to many months.
The X-ray spectrum of the persistent emission typically consists of a soft thermal component plus a hard power-law tail likely originating from multiple cyclotron resonant scattering in the magnetosphere \citep{Thompson_HEmagnetars_2005, baring_AXPs_2007}.
Persistent emission above $\SI{10}{\kilo\electronvolt}$ has been detected in a few magnetars, but upper limits (ULs) in the $\si{\mega\electronvolt}$ range indicate that this hard component cannot be extrapolated to higher energies \citep{denhartog_comptelUL_2006}.

Hard X-ray bursts have been detected from nearly all magnetars, typically during active periods when tens to hundreds of bursts are emitted within hours.
These bursts are emitted in the $\si{\kilo\electronvolt}-\si{\mega\electronvolt}$ bands, are short ($\sim \SI{0.1}{\second}$), have a total energy release of $\qtyrange{1E38}{1E40}{\erg}$, and peak X-ray luminosities of $\qtyrange{1E36}{1E43}{\erg\,\second^{-1}}$ \citep{israel_magnetarflares_2011, kozlova_sgr1935if_2016, kaspi_magnetars_2017}.
Some bursts, called `intermediate flares', are characterised by a longer duration (a few seconds to a few tens of seconds) and higher fluences \citep{olive_intermediateburst_2004}, while the much rarer magnetar `giant flares' \citep[MGF, ][]{mazets_gmf_1979, hurley_gmf_1999, palmer_gmf_2005, mereghetti_gmf_2024} reach peak X-ray luminosities of $\qtyrange{1E44}{1E47}{\erg\,\second^{-1}}$.

In the standard magnetar model, the magnetic field is the dominant source of energy, and it powers both the persistent and bursting emission \citep{israel_magnetarflares_2011}.
The persistent emission is primarily driven by the decay of the magnetic field, leading to the heating of the stellar crust and the emission of thermal radiation.
Non-thermal persistent emission can also arise from resonant cyclotron scattering in the magnetosphere \citep{mereghetti_magnetars_2015}.
Bursting emission occurs when magnetic stresses build up sufficiently to crack a patch of the NS crust, ejecting hot plasma into the magnetosphere \citep{thompson_outbursts_1995}.
Other models suggest that short bursts may arise from magnetic reconnection in the magnetosphere \citep{lyutikov_radiomagnetars_2002}.

In view of the rich phenomenology and the unique physical conditions found in these objects, the observation of magnetars in the gamma-ray range is of great interest.
However, only ULs have been derived at $\qtyrange{0.1}{10}{GeV}$ energies \citep{li_fermiulmagnetars_2017} when searching for the persistent emission.
Gamma-ray emission spatially coincident with some magnetars has been detected, but it is thought to originate from the surrounding supernova remnants \citep[SNRs,][]{li_fermiulmagnetars_2017}.
A candidate MGF was detected in the Sculptor galaxy NGC~253 at $\si{\giga\electronvolt}$ energies from $\qtyrange{19}{284}{\second}$ after the detection of an initial $\si{\mega\electronvolt}$ signal \citep{ajello_giant_flare_2021}.
The $\si{\giga\electronvolt}$ signal was likely generated by the interaction of an ultra-relativistic outflow of electrons (that first emitted the $\si{\mega\electronvolt}$ photons) with environmental gas.
The shock waves could accelerate the electrons to high energies and emit $\si{\giga\electronvolt}$ gamma rays as optically thin synchrotron radiation.
Imaging Atmospheric Cherenkov Telescopes (IACTs) performed intensive observation campaigns to monitor the very high-energy (VHE) emission from magnetars at $\si{\giga\electronvolt}-\si{\tera\electronvolt}$, without detecting a signal \citep{aleksic_observation_smagnetars_2013, hess_sgr1935_2021, magic_proceedingsgr_2021}.

The Soft Gamma Repeater (SGR)~1935$+$2154 was discovered in 2014 following the detection of a short burst and quickly recognised as a magnetar with spin period $P\sim\SI{3.24}{\second}$ and period derivative $\dot{P}=\SI{1.43(1)E-11}{\second \, \second^{-1}}$ \citep{israel_sgr1935discovery_2016}. 
These values correspond to a dipole magnetic field of $B_\text{dipole}\sim \SI{2.2E14}{\gauss}$, a characteristic age of $\sim \SI{3.6}{\kilo\year}$, and a spin-down power of $L_\text{sd}~\simeq~\SI{1.7E34}{\erg \, \second^{-1}}$.
SGR~1935$+$2154 is located close to the centre of SNR G57.2$+$0.8, which has an estimated distance of between $\qtyrange[range-phrase = \text{ and }]{6}{12}{\kilo\parsec}$ \citep{kothes_sgr1935snr_2018, zhou_revisiting_2020, lin_sgr1935properties_2020, lin_erratum_2020}. 
The analysis of expanding X-ray rings caused by dust scattering of bright bursts allowed \citet{mereghetti_integral_2020} to derive a magnetar distance of $4.4_{-1.3}^{+2.8}\,\si{\kilo\parsec}$, independent of its presumed association with the SNR.

This source has been the most prolific magnetar in recent years, with at least four active periods, including intermediate flares \citep{kozlova_sgr1935if_2016,lin_fermigbm_2020, denissenya_sgrtimeclustering_2021, borghese_sgr1935_2022, rehan_sgr1935_2023, ge_reanalysis_2023}.
The most notable event was a burst emitted on 28 April 2020, simultaneously detected in the radio \citep[FRB~20200428D, ][]{chimefrb_collaboration_bright_2020, bochenek_frb200428_2020} and X-ray bands \citep{mereghetti_integral_2020, ridnaia_peculiar_2021, hxmt_sgr_2021, tavani_x-ray_2021}.
The radio properties of this event are very similar to those of fast radio bursts (FRBs), millisecond-duration radio transients that originate from cosmological distances and whose origin is still unknown \citep{cordes_FRB_2019, petroff_FRB_2019}.
A wide variety of models have been proposed to explain the emission of FRBs, including magnetars, young isolated pulsars, mergers of compact objects, stellar-mass black holes, and cataclysmic events \citep{zhang_FRBphysics_2023}.
The discovery of FRB~20200428D, the first one detected from a source within the Galaxy and with a known origin, gave strong support to the classes of FRB models based on the presence of a magnetar and raised interest in multi-wavelength observations of SGR~1935$+$2154.
Many observations were carried out from the radio band \citep{bailes_sgr_multifrequency_2021} to IACTs such as H.E.S.S. and MAGIC, which performed observation campaigns at tera-electronvolt energies without detecting the magnetar \citep{hess_sgr1935_2021, magic_proceedingsgr_2021}.

Here we report on a search for tera-electronvolt emission from SGR~1935$+$2154 during periods of high activity of the source, using data taken with the Large-Sized Telescope prototype (LST$-$1) of the Cherenkov Telescope Array Observatory (CTAO) in 2021 and 2022 \citep{lst_performance_2023}.
In Sect.~\ref{sec:Observations} we describe the available observation dataset and Monte Carlo (MC) simulations.
In Sect.~\ref{sec:PersistentEmission} we describe the data analysis and results for the persistent emission, while in Sect.~\ref{sec:TransientEmission} we focus on the search for bursts. 
Finally, we discuss the constraints on the multi-wavelength emission of SGR~1935$+$2154 in Sect.~\ref{sec:discussion} and we conclude this work in Sect.~\ref{sec:conclusion} with a focus on future perspectives on magnetar observations and searches for FRB counterparts with IACTs.

\section{Observations and data processing}
\label{sec:Observations}
LST$-$1 observed SGR~1935$+$2154 for approximately $\SI{33}{\hour}$, over the course of $15$ nights in July 2021, September 2021, and June 2022.
The data were taken in `wobble mode', which allows for the simultaneous evaluation of the background \citep{fomin_falsesource_1994, berge_background_2007}.
The offset angle (i.e. the distance between the source position and the centre of the telescope's field of view), was $\approx \SI{0.4}{\deg}$ for each observation run.
The observations were taken up to $\SI{55}{\deg}$ zenith angle, and with different Moon illumination levels, but not every run was used as we applied quality selection criteria (see Sects.~\ref{sec:PersistentEmission}~and~\ref{sec:TransientEmission}).

The recorded data were processed using \verb|LSTOSA v0.9.2|, the semi-automatic pipeline of the LST Collaboration \citep{ruiz_lstosa_2022, morcuende_lstosa_2022}, which connects the different steps of \verb|lstchain|, the low-level analysis software developed for LST$-$1 \citep{lstchain_proceeding_2021, lstchain-Zenodo_2023}, based on \verb|ctapipe| \citep{ctapipe-icrc-2023}.
\verb|LSTOSA| acquires raw data from the camera (i.e. un-calibrated waveform signal), and performs calibration, charge integration in every pixel to produce Cherenkov images, image cleaning, and image parametrisation with Hillas parameters \citep{hillas_parameters_1985}.

The Hillas parameters are used to derive the physical properties of the incoming primary particles that generated the atmospheric shower: energy, arrival direction, and `gammaness'.
The gammaness parameter is a score between $0$ and $1$ that indicates how likely it is that the shower event was initiated by a gamma ray \citep{lst_performance_2023}.
We derived the event parameters with \verb|lstchain v0.9.13| by applying random forest (RF) algorithms \citep{breiman_randomforests_2001}, using the Hillas parameters as inputs.
The RFs were trained on MC simulated images of gamma-ray and hadronic events (protons), tuned to the night sky background (NSB) level of the data with the `noise padding' method \citep{lst_pevatron_2023,lst_performance_2023} using \verb|lstMCpipe v0.10.0| \citep{lstMCpipe_article_2022, lstMCpipe-Zenodo_2023}.
The spectrum of the simulated MC gamma-ray events is a power law with index $2$.
An independent set of MC simulated events was used to derive the `instrument response functions' (IRFs) after applying event selection cuts in Sects.~\ref{sec:PersistentEmission}~and~\ref{sec:TransientEmission}.

\section{Persistent emission data analysis}
\label{sec:PersistentEmission}
To evaluate the persistent emission of SGR~1935$+$2154, we selected a sub-sample of our dataset, considering only high-quality observations, i.e. runs in dark time (no Moon contamination) with a sufficiently high rate of cosmic events to avoid instrumental or environmental issues.
This selection resulted in $\SI{25.5}{\hour}$ of high-quality data, distributed over $13$ nights.

We produced the photon lists and the IRFs from the real and MC-simulated event lists, respectively, by applying event selection cuts that were optimised for a spectral analysis in \citet{lst_performance_2023}, and that can be applied to most standard spectral analyses performed with LST$-$1 \citep[e.g.][]{lst_pevatron_2023}.
The applied cuts are intensity, $i_\text{cut}=\SI{80}{\photoelectrons}$, energy-dependent `theta-containment', $\theta_\text{cont}=0.68$ (i.e. the point spread function containment fraction), and energy-dependent `gamma-hadron separation efficiency', $\epsilon_\text{gh}=0.70$.
We produced `point-like' IRFs \citep{nigro_gadf_2021} valid for a fixed offset value of $\SI{0.4}{\deg}$.

\begin{figure}[ht!]
   \centering
   \resizebox{\hsize}{!}{\includegraphics{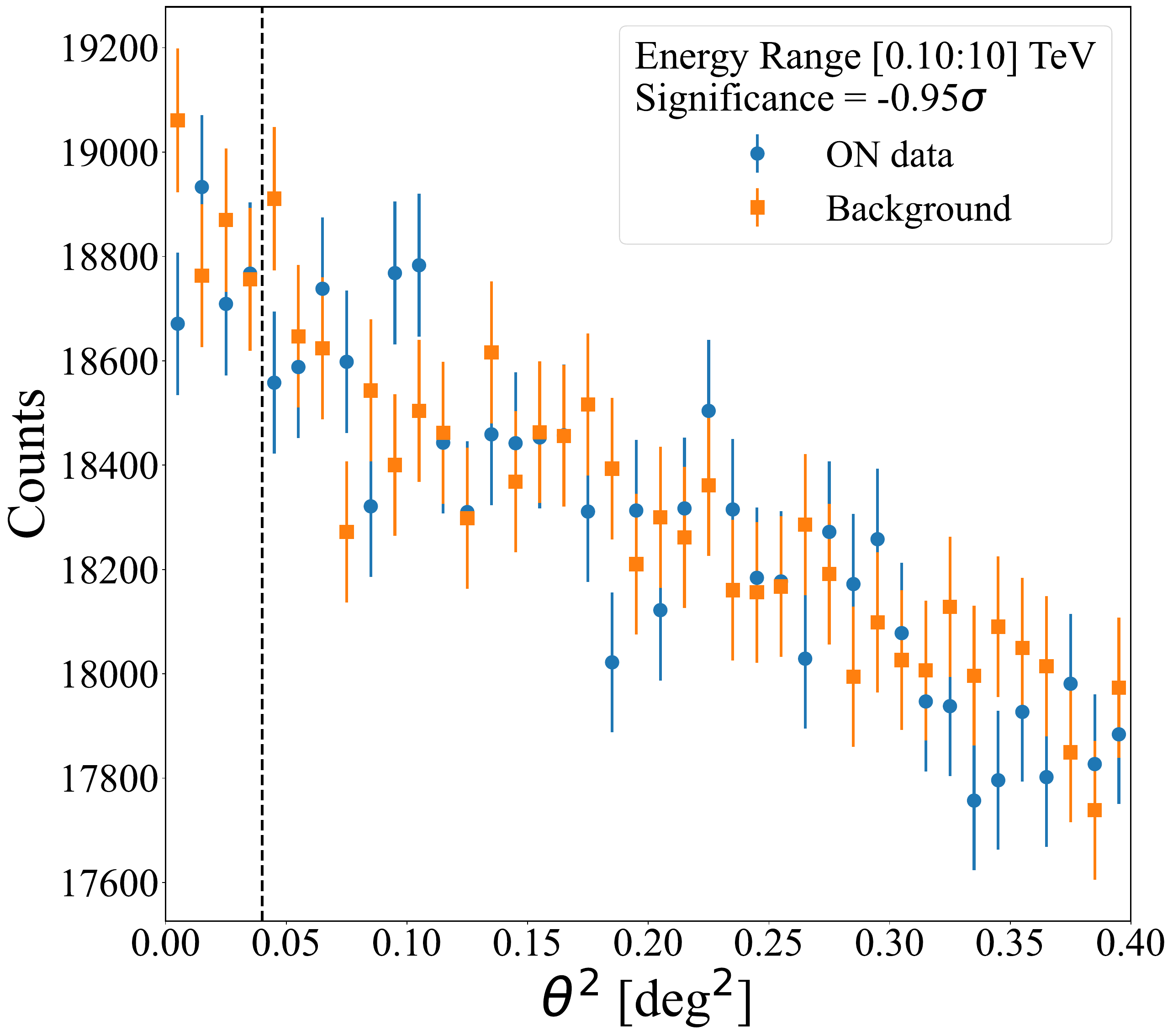}}
   \caption{$\theta^2$ plot on SGR~1935$+$2154 persistent emission.
   We used $\SI{25.5}{\hour}$ of high-quality data (see Sect.~\ref{sec:PersistentEmission}), which we show here in a single energy bin from $\qtyrange{0.1}{10}{\tera\electronvolt}$.
   We show the distribution of ON counts in blue and the background distribution in orange. The dashed line represents the $\theta^2$ cut used to evaluate the significance.
   No emission is detected.
 }
 \label{fig:PersistentEmissionTheta2}
\end{figure}

We performed the high-level analysis with \verb|gammapy| \verb|v1.3| \citep{gammapy_paper_2023, gammapy_software1.3_2025} by stacking all data of the selected runs.
The energy thresholds of the observations exhibit a wide distribution due to their dependence on the zenith angle.
We used a conservative minimum analysis energy of $\SI{0.1}{\tera\electronvolt}$ to ensure that the telescope's effective area remains above $10\%$ of its maximum value across all runs.
We estimated the number of background events in the signal region through the `reflected regions' method \citep{berge_background_2007}, using one OFF region with a size equal to that of the ON region, defined by the $\theta_\text{cont}$ selection cut, energy-dependent radii typically in the $\qtyrange{0.12}{0.22}{\deg}$ range.
The $\theta^2$ plot for the selected data does not show any significant signal (see Fig.~\ref{fig:PersistentEmissionTheta2}).

\begin{figure}[ht!]
   \centering
   \resizebox{\hsize}{!}{\includegraphics{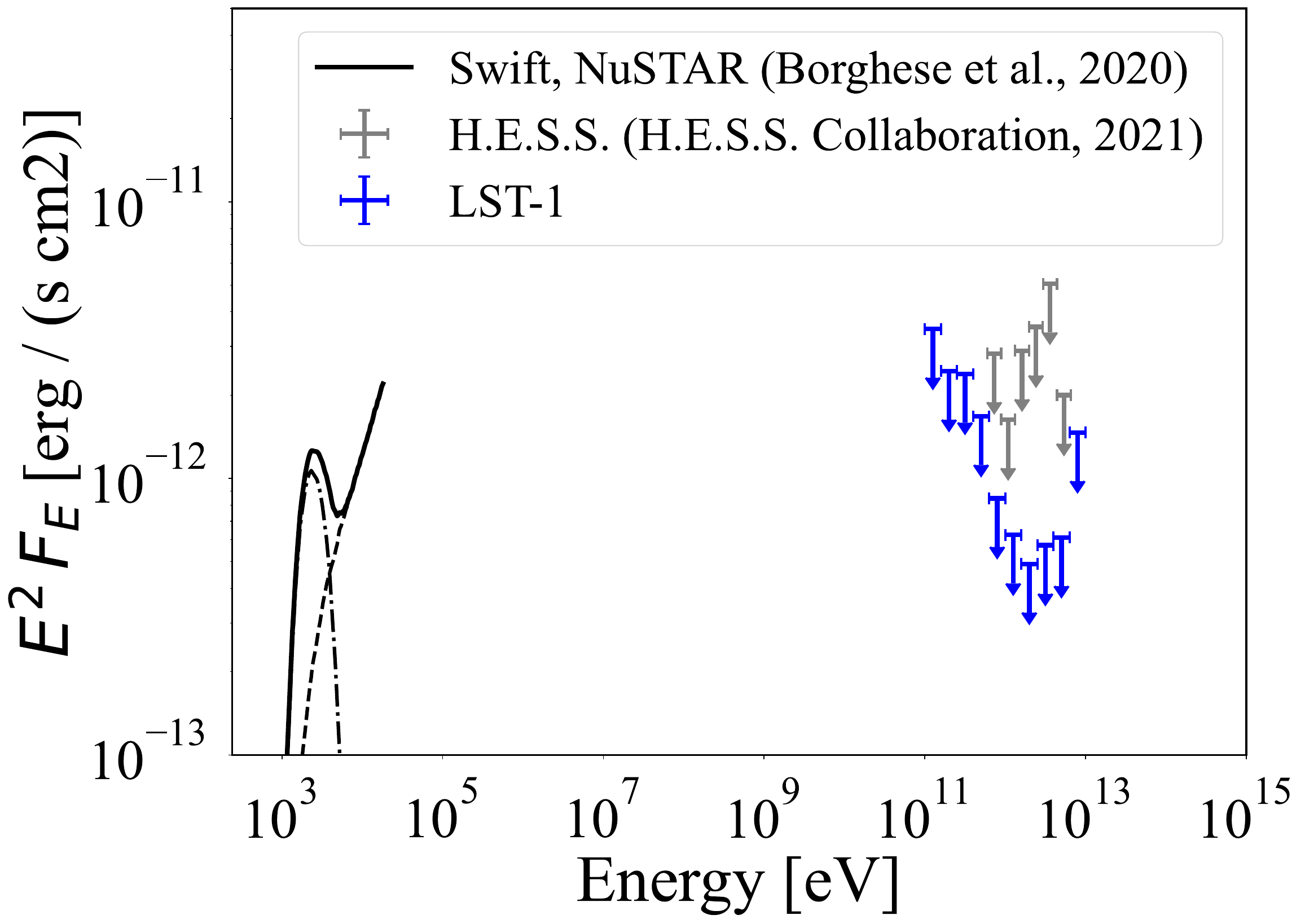}}
   \caption{Multi-band SED of the persistent emission of SGR~1935$+$2154.
   Black lines show the best fit emission model in the X-ray and soft gamma-ray bands \citep{borghese_sgr1935_2022}.
   In the VHE band, our $95\%$ confidence level ULs confirm the non-detection obtained by \citet{hess_sgr1935_2021}.
   The tera-electronvolt ULs are about the same order of magnitude as the X-ray emission and confirm previous studies, which suggest that the power-law component observed above $>\SI{10}{\kilo\electronvolt}$ must have a break at $\si{\mega\electronvolt}$ energies.
   }
 \label{fig:PersistentEmissionSED_E2DNDE}
\end{figure}

As we did not detect the source, we performed a $1$D spectral analysis with the reflected-regions background method to estimate the $95\%$ confidence level ULs, which we show in Fig.~\ref{fig:PersistentEmissionSED_E2DNDE}.
We used ten logarithmically spaced bins between $\SI{0.1}{\tera\electronvolt}$ and $\SI{10}{\tera\electronvolt}$ and assumed a point-like source with a power law spectrum with photon index $2.5$, an average value for tera-electronvolt sources that have not a known emission model in literature, already proposed by \citet{hess_sgr1935_2021}.
The UL on the integrated flux in the $\qtyrange{0.1}{10}{\tera\electronvolt}$ range is $\SI{5.6E-12}{\second^{-1}\centi\meter^{-2}}$, corresponding to $\SI{2.4E-12}{\erg\,\second^{-1}\centi\meter^{-2}}$.
The flux UL above $\SI{0.6}{\tera\electronvolt}$, for direct comparison with \citet{hess_sgr1935_2021}, is $\SI{3.0E-13}{\second^{-1}\centi\meter^{-2}}$ (corresponding to $\SI{6.8E-13}{\erg\,\second^{-1}\centi\meter^{-2}}$).
We also searched for source variability by computing the nightly light curve of the emission and the nightly spectral energy distribution (SED, $\approx \SI{2}{\hour}$ livetime per night), but we did not detect the source in any of the observation nights.

\begin{figure*}[ht!]
   \centering
   \resizebox{\hsize}{!}{\includegraphics{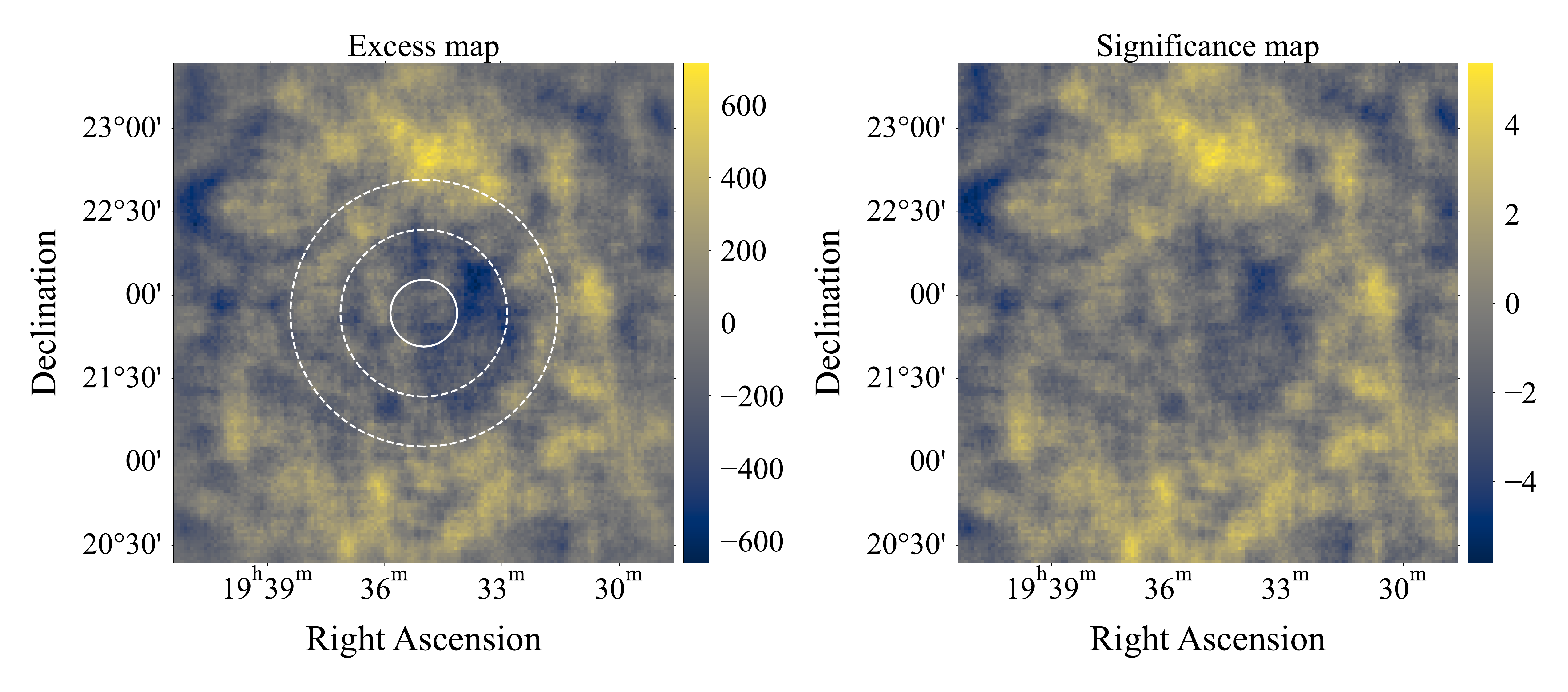}}
   \caption{
   Excess (left) and statistical significance (right) maps in a $\qtyproduct{3 x 3}{\deg}$ region centred on SGR~1935$+$2154 in the $\qtyrange[range-phrase = \text{ and }]{0.1}{10}{\tera\electronvolt}$ energy range.
   The ON region is shown as a white circle, while the background ring is delimited by the two dashed circles.
   The maps do not show any significant excess on the source.
   }
 \label{fig:PersistentEmissionMaps}
\end{figure*}

For visualisation purposes we produced the excess and significance $2$D maps centred on SGR~1935$+$2154, using the ring-background model in \verb|gammapy|.
Since the ring spans regions with offsets different from the ON region, the detector acceptance cannot be assumed to be constant, and its profile must therefore be computed explicitly \citep{berge_background_2007}.
We evaluated the radial acceptance from the data on a run-by-run basis and stacked the results to estimate the background for the entire dataset \citep{lst_pevatron_2023}, using the \verb|BAccMod|\footnote{\url{https://github.com/mdebony/BAccMod}} package.
The position of SGR~1935$+$2154 was masked to prevent contamination.
We defined the OFF region as a ring with internal radius of $\SI{0.5}{\deg}$ and $\SI{0.3}{\deg}$ width, around a $\SI{0.2}{\deg}$ radius circular ON region.
The maps, shown in Fig.~\ref{fig:PersistentEmissionMaps}, do not show any significant excess on the source.

\section{Bursting emission data analysis}
\label{sec:TransientEmission}

To search for VHE bursting emission from SGR~1935$+$2154, we extracted a new dataset of photon lists and IRFs using a set of event selection cuts optimised to detect a gamma-ray signal on a $\SI{0.1}{\second}$ timescale (typical duration of a short magnetar X-ray burst), in a low-photon-statistics regime and under the hypothesis of a Poisson background \citep{cowan_smallNinHEP_2007, patrignani_reviewparticlephysics_2016, magic_frb_2018}.
We used the following `burst cuts': a cut in gammaness, $g_{cut}=\num{0.75}$, in intensity, $i_{cut}=\SI{50}{\photoelectrons}$, and in $\theta^2_{cut}=\SI{0.08}{\deg^2}$.

The LST$-$1 observations included nine hard X-ray bursts from SGR~1935$+$2154 detected by various high-energy (HE) space telescopes (see Table~\ref{table:BurstUpperLimits}).
We searched for VHE counterpart emission simultaneous to these events, and also performed an unbiased search for short bursts in the whole dataset.

\subsection{Search for tera-electronvolt emission simultaneous to X-ray bursts} \label{sec:TransientEmission:sub:ToA}
We selected the nine LST$-$1 runs comprising the time of arrival (ToA) listed in Table~\ref{table:BurstUpperLimits}, and the point-like IRFs for the closest simulation node to every run.
We used the IRFs to obtain the exposure, assuming a power-law spectrum with index $2$.

\begin{table*}[ht!]
 \centering
 \caption{
  Upper limits in the $\qtyrange[range-phrase=-, range-units=bracket, range-open-bracket=[,range-close-bracket=]]{0.1}{10}{\tera\electronvolt}$ range on known hard X-ray from SGR~1935$+$2154 observed by different satellites.
  \label{table:BurstUpperLimits}
 }
 \begin{tabular}{l l l | c | c c | c | c c c }
      \hline\hline
      $\#$ & Time of Arrival & Instrument & Exposure & $R_\text{bkg}$ & $N_{5\sigma}$ & $N_\text{ON}$ & $s_\text{UL}$ & Flux UL & Fluence UL \\
         & ISOT UTC & &  $\SI{1E8}{\centi\meter^2\second}$ & $\si{\second^{-1}}$ & & & & $\SI{1E-8}{\centi\meter^{-2}\second^{-1}}$ & $\SI{1E-9}{\erg \,\centi\meter^{-2}}$ \\
      \hline
      1 & 2021-07-07 00:33:31.670 & \textit{Fermi}-GBM & 1.448 & 0.81$\pm$0.02 & 4 & 0 & 2.16 & 1.49 & 1.09 \\
      2 & 2021-09-10 23:40:34.460 & \textit{Fermi}-GBM & 1.485 & 1.05$\pm$0.03 & 4 & 0 & 2.19 & 1.47 & 1.08 \\
      3 & 2021-09-11 22:51:41.600 & \textit{GECAM}     & 1.430 & 0.95$\pm$0.03 & 4 & 0 & 2.17 & 1.52 & 1.11 \\
      4 & 2021-09-11 23:55:45.872 & \textit{NICER}     & 1.485 & 1.01$\pm$0.03 & 4 & 0 & 2.18 & 1.47 & 1.07 \\
      5 & 2021-09-12 00:34:37.450 & \textit{GECAM}     & 1.491 & 0.61$\pm$0.03 & 4 & 0 & 2.13 & 1.43 & 1.04 \\
      6 & 2021-09-12 00:45:49.400 & \textit{GECAM}     & 1.491 & 0.66$\pm$0.03 & 4 & 0 & 2.14 & 1.43 & 1.05 \\
      7 & 2021-09-12 22:16:36.200 & \textit{GECAM}     & 1.296 & 0.68$\pm$0.02 & 4 & 1 & 3.86 & 2.97 & 2.17 \\
      8 & 2021-09-12 23:19:32.080 & \textit{Fermi}-GBM & 1.430 & 1.04$\pm$0.03 & 4 & 0 & 2.18 & 1.53 & 1.12 \\
      9 & 2021-09-13 00:27:25.200 & \textit{GECAM}     & 1.485 & 1.04$\pm$0.03 & 4 & 0 & 2.18 & 1.47 & 1.07 \\
      \hline
      & stacked $\delta t=\SI{0.9}{\second}$ &        & 13.041 & 0.87$\pm$0.04 & 8 & 1 & 3.57 & 0.27 & 0.20 \\ 
      \hline
 \end{tabular}
 \tablefoot{The exposure is given for a $\delta t=\SI{0.1}{\second}$ observation time assuming a power-law spectrum with index $2$.
  The background rate, $R_\text{bkg}$, was estimated in a OFF region and was used to estimate $N_{5\sigma}$, the counts needed for a $5\sigma$ detection.
  The $N_\text{ON}$ counts were evaluated in a $\SI{0.1}{\second}$ time window centred around the burst ToA and were used to compute $s_\text{UL}$ according to eq.~\eqref{eq:bayesianupperlimitpoisson}.
  The photon flux ULs were given by dividing $s_\text{UL}$ over the exposure.
  The fluence ULs were obtained converting the photon flux into energy flux and integrating over $\delta t$.
  The stacked entry assumes an average $R_\text{bkg}$ and effective area.
  Burst $\#1$ was detected by \citet{atel_14764_2021, gcn_30407_2021,gcn_30400_2021,gcn_30458_2021,gcn_30418_2021}.
  $\#2$ by \citet{gcn_30806_2021, atel_14907_2021}.
  $\#3$, $\#5$ and $\#6$ by \citet{gcn_30822_2021}.
  $\#4$ by \citet{atel_14916_2021}.
  $\#8$ by \citet{gcn_30831_2021}.
  $\#7$ and $\#9$ by \citet{gcn_30836_2021}.}
\end{table*}

For each of the nine bursts listed in Table~\ref{table:BurstUpperLimits}, we selected the number of photons, $N_\text{ON}$, in the ON region centred on SGR~1935$+$2154 (with size given by $\theta_\text{cut}$), in a $\delta t=\SI{0.1}{\second}$ time window centred around the ToA of each of the nine bursts, and in the $\qtyrange{0.1}{10}{\tera\electronvolt}$ energy range.
No timing correction was applied to the ToAs, which represent the detection times of the X-ray bursts.  
This choice is justified because all instruments in Table~\ref{table:BurstUpperLimits} are in low Earth orbit at altitudes of $\lesssim \SI{600}{\kilo\meter}$.  
The resulting time delay of the signal with respect to the LST$-$1 site is $\lesssim \SI{2}{\milli\second}$, which is negligible compared to the $\SI{0.1}{\second}$ analysis window.

None of the nine bursts is detected by LST$-$1, as $N_\text{ON}$ is always lower than the $N_{5\sigma}$ threshold needed to claim a detection.
The threshold depends on the measured background rate, $R_\text{bkg}$.
The background rates of the nine selected observation runs are reported in Table~\ref{table:BurstUpperLimits}, with the associated Poisson error.

The Bayesian treatment of the ON/OFF problem has been extensively discussed in the literature \citep[e.g.][]{knoetig_bayesianonoff_2014,casadei_bayesiananalysisonoff_2015}.
The number of counts in the ON region can be modelled with a Poisson likelihood, parametrised by the expected signal, $s$, and background, $b$, counts:
\begin{equation}\label{eq:Casadei_Likelihood_full}
    \Li (n|s,b) = \Poi (n;s+b)= \frac{(s+b)^n}{n!} \, e^{-(s+b)},
\end{equation}
where $\Poi$ denotes the Poisson distribution, and $n$ is the statistical variable describing the ON counts (while we used $N_\text{ON}$ for the actually measured values).
Assuming statistical independence between $s$ and $b$, the joint prior factorises as the product of the signal prior, $\pi_\text{sig}(s)$, and the background prior, $\pi_\text{bkg}(b)$.
Once the priors have been derived, the UL to the signal parameter, $s_\text{UL}$, can then be derived in three steps.
First, the likelihood is marginalised over the background parameter: 
\begin{equation}
\label{eq:Casadei_Likelihood_marginalized}
   \Li_\text{sig}(n|s) = \int_0^\infty \, \pi_\text{bkg} (b)\, \Li(n|s,b) \, \d b.
\end{equation}
Then, the marginalised posterior distribution is obtained:
\begin{equation}\label{eq:Casadei_Posterior}
    \Po_\text{sig}(s|n) = \frac{\pi_\text{sig}(s) \cdot \Li_\text{sig}(n|s)}{\int_0^\infty \, \pi_\text{sig} (x)\, \Li_\text{sig}(n|x) \, \d x}.
\end{equation}
Finally, the UL $s_\text{UL}$ at confidence level 
$\alpha_\text{CL}$ is defined as the solution of
\begin{equation}\label{eq:BayesianUpperLimit}
    1-\alpha_\text{CL} = \int_0^{s_\text{UL}} \, \Po_\text{sig}(s|n) \, \d s.
\end{equation}
The choice of priors for $s$ and $b$ is crucial in this approach and will now be examined.
A common, though not always well-justified, approach is to assume a Dirac delta function for the background prior:
\begin{equation}
\label{eq:BackgroundPriorDiracDelta}
    \pi_\text{bkg}(b)=\delta(b-b_0),
\end{equation}
where $b_0$ is the expected background counts, usually estimated from the OFF region as $\displaystyle b_0 = \alpha \cdot N_\text{OFF}$, with $\alpha$ being the exposure ratio and $N_\text{OFF}$ the observed counts in the OFF region.
A more general and statistically consistent approach, proposed by \citet{casadei_bayesiananalysisonoff_2015}, derives the background prior directly from the posterior distribution of the OFF region.
The OFF region is described by a single Poisson likelihood, $\displaystyle \Li_\text{OFF}(k|B)=\Poi(k;B)$, where $k$ is the statistical variable and $B$ the Poisson parameter.
The prior for the OFF region can be naturally modelled as a Gamma distribution:
\begin{equation}
\label{eq:GammaPDF}
    \pi_\text{OFF} (B)= f_\Gamma \left(B; \nu, \rho \right) = \frac{\rho^\nu}{\Gamma \left(\nu \right)} \cdot B^{\nu-1} \cdot e^{-\rho B}
    \,,\quad B>0, \, \nu>0, \, \rho>0,
\end{equation}
where $\nu$ is the shape parameter, $\rho$ the rate, and $\Gamma$ the complete Gamma function.
Since the Gamma distribution is conjugate to the Poisson model, the posterior for the OFF region is also Gamma-distributed, reading as
\begin{equation}
\label{eq:OFF_Posterior}
    \Po_\text{OFF} (B|k) = f_\Gamma \left(B; \nu+k, \rho+1 \right).
\end{equation}
By rescaling from the OFF to the ON region via the exposure ratio ($\displaystyle b= \alpha \cdot B$), the prior for the ON-region background becomes
\begin{equation}
\label{eq:Prior_Background}
    \pi_\text{bkg} (b) = f_\Gamma \left(b; \nu+k, \frac{\rho+1}{\alpha} \right).
\end{equation}
A well-known property of the Gamma distribution is that when both the shape and rate parameters tend to infinity, while their ratio $b_0$ remains constant, the distribution converges to a Dirac delta function centred at $b_0$.
In eq.~\eqref{eq:Prior_Background}, this ratio is given by
\begin{equation}
\label{eq:GammaDeltaLimit}
    b_0 = \frac{\alpha\,(\nu+k)}{\rho+1}.
\end{equation}
In general, the choice of the shape and rate parameters define how the background is characterised.
For example, adopting Jeffreys' prior for $\pi_\text{OFF}(B)$ corresponds to setting $\nu = 1/2$ and $\rho = 0$ \citep{casadei_bayesiananalysisonoff_2015}.
Since the best estimator for the expected value of $k$ is the observed number of counts in the OFF region, $N_\text{OFF}$, the shape-to-rate ratio can be estimated as $\displaystyle b_0 = \alpha \cdot \left(N_\text{OFF}+1/2\right)$.

If the selected OFF region satisfies $\alpha \ll 1$ and $N_\text{OFF} \gg 1$, the background prior, $\pi_\text{bkg}$, in eq.\eqref{eq:Prior_Background} tends to a Dirac delta with $b_0 \approx \alpha \cdot N_\text{OFF}$ (neglecting the $1/2$ term), as in eq.\eqref{eq:ON_Prior_Dirac}.
Within this framework, and under these specific conditions, the use of a Dirac delta as background prior is not merely an ad hoc assumption, but can be justified as the approximation of a `well-characterised background'.

For our analysis, we selected a single OFF region, symmetrical to the ON region with respect to the pointing direction and of equal size.
The exposure ratio is therefore given by the ratio of the observing times, $\displaystyle \alpha=\frac{ T_\text{ON}}{ T_\text{OFF}}$, where $T_\text{ON} = \SI{0.1}{\second}$ and $T_\text{OFF}$ corresponds to the full duration of the observation run ($\approx 15-20$ $\si{\minute}$).
This yields $\alpha \sim 10^{-4}$, with measured OFF counts in the range of $N_\text{OFF} \approx 500$–$1000$ across the selected runs.

Our observations satisfy the conditions for the well-characterised background approximation, provided that the event rate remains constant, ensuring that the background can be described by a single Poisson distribution.
In practice, the event rate is not perfectly constant due to variations in atmospheric conditions and telescope pointing during a run.
To account for this, we tested all selected observations to verify that the distribution of binned OFF counts is consistent with a single Poisson law using a $\chi^2$ test, and rate variations are negligible.

\begin{figure}[ht!]
   \centering
   \resizebox{\hsize}{!}{\includegraphics{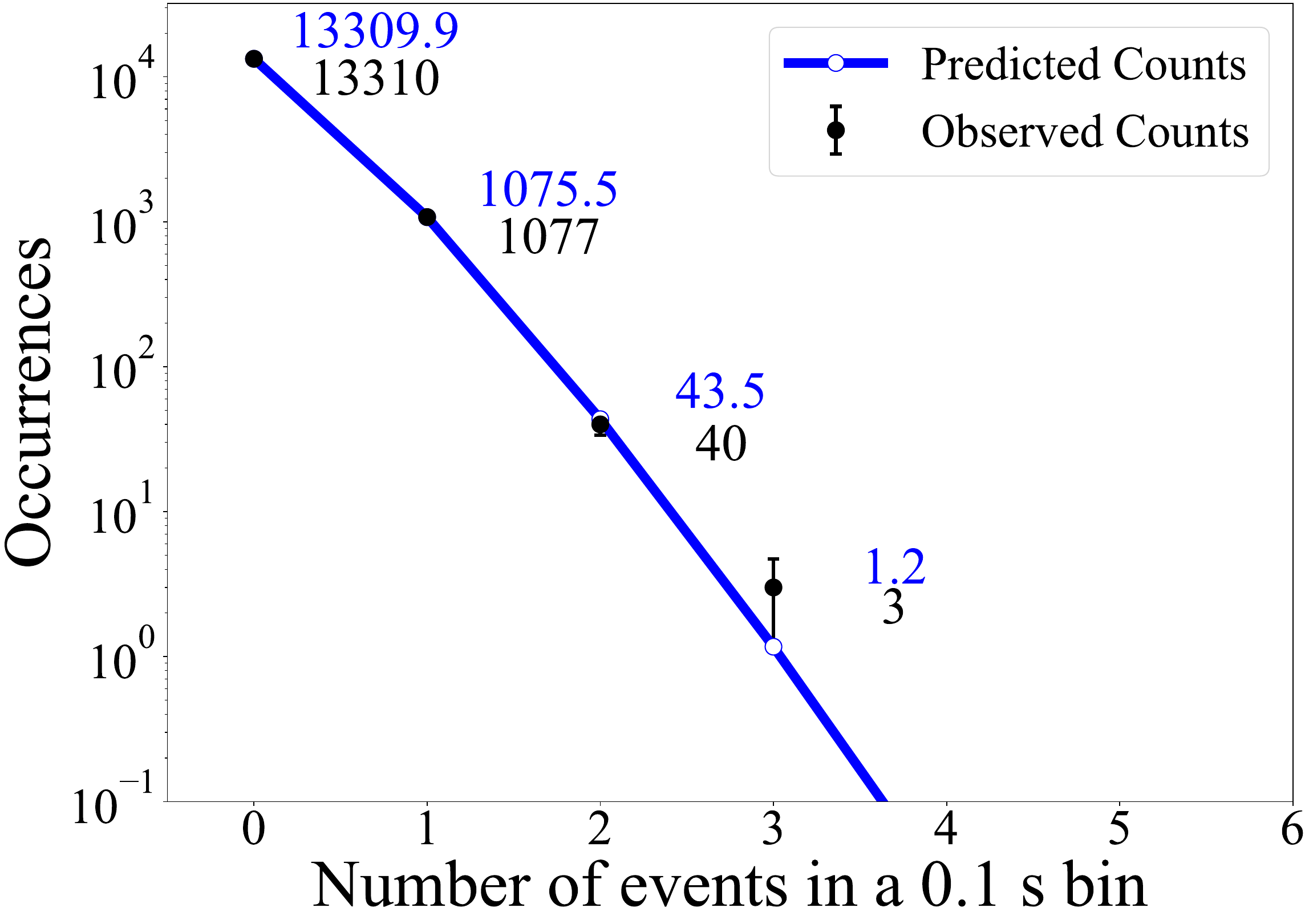}}
   \caption{Distribution of binned OFF counts (black points and labels) in the observation run of burst $\#1$ of Table~\ref{table:BurstUpperLimits}.
   The data are consistent with the expectation from a Poisson distribution (blue points, labels and line) with background rate $R_\text{BKG}=\SI{0.81}{\second^{-1}}$, using bins of width $\delta t=\SI{0.1}{\second}$.
   The total duration of the run was $\SI{1443}{\second}$.
   }
 \label{fig:Background}
\end{figure}

As an example, Fig.~\ref{fig:Background} shows the distribution of binned OFF counts for the observation run of burst $\#1$ in Table~\ref{table:BurstUpperLimits}.
The observed counts are consistent with the Poisson prediction, yielding a $\chi^2$ statistic of $3.1$ for $3$ degrees of freedom, corresponding to a $p$ value of $0.3$. Similar results for the other observation runs support the validity of the well-characterised background approximation.
Under these conditions, we consider adopting eq.~\eqref{eq:BackgroundPriorDiracDelta} as the background prior to be justified.

Once the background prior is fixed, the marginalised likelihood, $\Li_\text{sig}(n \mid s)$, can be computed as in eq.~\eqref{eq:Casadei_Likelihood_marginalized}.
Under the well-characterised background approximation, this reduces to the offset Poisson distribution $\displaystyle \Poi(n;s+b_0)$, whereas in the general case it is proportional to an exponential times a polynomial \citep[for a detailed discussion, see][]{casadei_referenceanalysis_2012, casadei_referenceprior_2014}.

Having established the treatment of the background prior and the resulting marginalised likelihood, we now turn to the choice of the signal prior, $\pi_\text{sig}(s)$.
A common assumption is to adopt a uniform prior for $s > 0$, mainly for its mathematical simplicity \citep{cowan_smallNinHEP_2007, patrignani_reviewparticlephysics_2016, magic_frb_2018}, and because it is often considered as a `non-informative' prior.
In reality, it does not result from a formal procedure for constructing an objective prior .
Moreover, it disproportionately favours large values of the signal, even nonphysical large ones, which leads to an overestimation of both the posterior mean and the corresponding ULs \citep{casadei_referenceprior_2014}.

By contrast, Jeffreys' prior is derived solely from the Fisher information of the likelihood and can therefore be regarded as an objective prior.
For an offset Poisson distribution, such as the marginalised likelihood $\displaystyle \Li_\text{sig}(n|s)=\Poi(n;s+b_0)$, it is
\begin{equation}
\label{eq:ON_Prior_Dirac}
    \pi_\text{sig}(s) = \sqrt{\frac{b_0}{s+b_0}},
\end{equation}
which we adopted as our signal prior.
It is important to note that Jeffreys' prior depends on the marginalised likelihood, and therefore implicitly carries information from the background prior.
If $\pi_\text{bkg}(b)$ is not a Dirac delta, the signal prior cannot be computed analytically. Eq.~\eqref{eq:ON_Prior_Dirac} should thus be regarded as the approximation of the general prior under the well-characterised background assumption \citep[for a comparison of numerical and approximate priors, see][]{casadei_referenceprior_2014}.

Once the signal prior and the marginalised likelihood are specified, the posterior distribution can be obtained by multiplication and normalisation, as in eq.~\eqref{eq:Casadei_Posterior}.
The calculations are presented in Appendix~\ref{sec:appendix_integrals} for completeness.
Its explicit form is
\begin{equation}
    \label{eq:Casadei_Posterior_Dirac}
    \Po_\text{sig}(s|n) = \frac{ (s+b_0)^{n-\frac{1}{2}} \, e^{-(s+b_0)} }{ \Gamma\left(n+\frac{1}{2}\right) \cdot \left[ 1 - F_{\chi^2} \left( 2b_0; 2 n+1 \right) \right]},
\end{equation}
where $F_{\chi^2}$ is the cumulative distribution function of the $\chi^2$ distribution with $2 n+1$ degrees of freedom, and $b_0$ the expected background counts.
The UL on the expected signal counts, $s_\text{UL}$, follows from eq.~\eqref{eq:BayesianUpperLimit}:
\begin{equation}
\label{eq:bayesianupperlimitpoisson}
    \begin{cases}
    s_\text{UL} = \frac{1}{2} \, F^{-1}_{\chi^2} \left(p; 2 n+1 \right)-b_0,\\
    p = 1 -\alpha_\text{CL} \cdot \left( 1 - F_{\chi^2} \left( 2 b_0; 2 n+1 \right) \right)
    \end{cases}
,\end{equation}
where $F^{-1}_{\chi^2}$ is the inverse cumulative distribution (percent-point) function of the $\chi^2$ distribution with $2 n+1$ degrees of freedom, $b_0$ the expected background counts, and $1-\alpha_\text{CL}$ is the confidence level.
The calculations are presented in Appendix~\ref{sec:appendix_integrals} for completeness.
The choice of prior directly affects the degrees of freedom of the $\chi^2$ distribution: for Jeffreys' prior the term is $2n+1$, whereas for a uniform prior it would be $2(n+1)$.
Consequently, ULs computed with the uniform prior are systematically larger than those obtained with Jeffreys' prior, reflecting the fact that Jeffreys' prior assigns monotonically decreasing weight to higher signal intensities.

We report the $95\%$ confidence level ULs $s_\text{UL}$ for every burst in Table~\ref{table:BurstUpperLimits}.
The estimation of the UL in practice was performed through eq.~\eqref{eq:bayesianupperlimitpoisson}, using the measured counts in the ON region, $N_\text{ON}$, as an estimator for $n$, and $b_0$ as the background rate, $R_\text{bkg}$, times $\SI{0.1}{\second}$.
We computed the corresponding ULs on the integral flux by dividing $s_\text{UL}$ over the exposure of the time window.
The average photon flux UL for a $\SI{0.1}{\second}$ burst is $\SI[separate-uncertainty-units = bracket]{1.6(5)E-8}{\centi\meter^{-2} \second^{-1}}$ and the average energy fluence UL is $\SI[separate-uncertainty-units = bracket]{1.2(4)E-9}{\erg\,\centi\meter^{-2}}$.

We stacked the counts and the exposure of every burst window to compute the ULs on the total VHE burst activity of SGR~1935$+$2154, reported in Table~\ref{table:BurstUpperLimits} under the assumption of average background and effective area.
The stacked photon flux UL on the bursting emission is then $\SI{2.7E-9}{\centi\meter^{-2}\second^{-1}}$.

\subsection{Non-simultaneous burst search} 
\label{sec:TransientEmission:sub:BurstSearch}
We searched for possible SGR~1935$+$2154 short bursts in our dataset.
We binned the ON-source events in $\SI{0.1}{\second}$ bins starting at the beginning of each observation run.
We evaluated if any of the bins showed a post-trial significance of five standard deviations \citep{li_ma_analysis_1983, bulgarelli_mle_2012, brun_analysis_2020}, and we did not find any detection.

The mean and standard deviation of the background rates across the runs of our dataset are $R_\text{bkg}=\SI{0.9(3)}{\second^{-1}}$, which provide pre-trial $N_{5\sigma}$ from $\numrange{3}{5}$.
Differences in the runs can be attributed to different zenith angle, atmospheric conditions, and NSB levels.
We performed the burst search across all our dataset, considering each $\SI{0.1}{\second}$ time bin as a trial, with the total number of trials $N_\text{trials}=\num{1355170}$.
The corrected post-trial $\tilde{N}_{5\sigma}$ range from $\numrange{5}{8}$.
The flux sensitivity for our non-simultaneous burst search can be evaluated by dividing the highest $\tilde{N}_{5\sigma}=8$ over the lowest $\SI{0.1}{\second}$ exposure of our dataset, which is $\SI{1.296E8}{\centi\meter^2\second}$, providing a minimum detectable photon flux of $\SI{6.2E-8}{\centi\meter^{-2}\second^{-1}}$ in the $\qtyrange{0.1}{10}{\tera\electronvolt}$.
Its value is higher than that of the analysis described in Sect.~\ref{sec:TransientEmission:sub:ToA} due to the large number of trials.

\section{Discussion}
\label{sec:discussion}
We analysed over $\SI{25}{\hour}$ of VHE gamma-ray observations of SGR~1935$+$2154 with the LST$-$1 during periods of known magnetar activity.
We did not detect persistent or bursting emission from the source but we provide ULs that constrain its emission in the
$\qtyrange{0.1}{10}{\tera\electronvolt}$ energy range.
We discuss below our constrains in the context of multi-wavelength observations of SGR~1935$+$2154 and its possible emission mechanisms.

\subsection{Persistent emission}
\label{sec:discussion:sub:persistent}
As is seen in Sect.~\ref{sec:PersistentEmission}, no persistent VHE signal is detected from SGR~1935$+$2154.
In Fig.~\ref{fig:PersistentEmissionSED_E2DNDE}, we present the multi-band SED of SGR~1935$+$2154.
Black lines show the best fit emission model in the X-ray and soft gamma-ray bands extracted from June 2020 observations of the source with \textit{Swift} and \textit{NuSTAR}.
The model is composed of a black-body spectrum plus a power-law component \citep{borghese_sgr1935_2022}.
In the VHE band, our ULs improve those by \citet{hess_sgr1935_2021}, extending the covered energy range down to $\SI{0.1}{\tera\electronvolt}$ and lowering them by factors of $\numrange{2}{5}$.
The tera-electronvolt ULs are about the same order of magnitude of the X-ray emission of SGR~1935$+$2154, and are consistent with previous studies on the source, which suggest that the power-law component observed in the SED above $>\SI{10}{\kilo\electronvolt}$ must have a break at $\sim \si{\mega\electronvolt}$ energies.

Most models explain the persistent emission of magnetars in the X-ray band using two components.
The first one is the thermal emission arising from the NS surface and its thin atmosphere \citep{turolla_magnetars_2015}.
The second one is due to the presence of accelerated particles in a magnetosphere with a complex geometry, which can modify the emitted spectrum producing hard tails by resonant cyclotron scattering of the thermal photons \citep{thompson_magnetarpersistent_2002}.
More complex models include MC simulations to study the motions of charged particles flowing in twisted magnetic loops \citep[e.g.][]{Fernandez_magnetarmodels_2007,beloborodov_magnetarpersistent_2013}.
Current models can produce power-law emission spectra with a suppression above the $\si{\mega\electronvolt}$ range, but the underlying mechanisms are not fully understood yet, mainly due to the lack of data in this energy range.
The cut-off has not been observed yet, and ULs are of the order of $\sim\SI{E-10}{\erg\per\second\per\centi\meter\squared}$ in the $\si{\mega\electronvolt}$ range \citep[e.g. ][]{denhartog_comptelUL_2006} and $\sim\SI{E-12}{\erg\per\second\per\centi\meter\squared}$ in the $\si{\giga\electronvolt}$ range \citep{li_fermiulmagnetars_2017}.
Future facilities capable of observing at $\sim \SI{1}{\mega\electronvolt}$, such as the upcoming COSI satellite mission \citep{tomsick_COSI_2023}, may detect the power law cut-off and validate these models.

Magnetars might also be a possible component of the progenitors of FRBs \citep{cordes_FRB_2019, zhang_FRBphysics_2023}.
For instance, FRB~121102 \citep{spitler_FRB121102_2014} is associated with a persistent radio source \citep[PRS, ][]{chatterjee_FRB121102localization_2017} and an optical counterpart classified as a low-metallicity star-forming dwarf galaxy \citep{tendulkar_FRB121102host_2017, bassa_FRB121102SFregion_2017}.
The PRS emission is not identified, but is likely due to synchrotron emission of relativistic electrons from a surrounding nebula \citep[e.g. ][]{murase_burst_2016, marcote_frb121102models_2017, metzger_frb121102model_2017, Bhattacharya_PRS_2024}, possibly a SNR, a pulsar wind nebula (PWN), or a hyper-accreting X-ray binary \citep{sridhar_FRB_binaries_2022}.
The recent detection of a PRS for FRB~20190520B \citep{Niu_PRS_FRB20190520B_2022}, FRB~20201124A \citep{bruni_PRS_FRB20201124A_2024}, and FRB~20240114A \citep{bruni_Atel16885_PRS_FRB20240114A_2024} supports nebular models powered by a NS as being the origin of FRBs.
Even though no HE counterparts of FRBs or their associated PRS have been detected \citep{Scholz_FRB121102mwl_2017, zhang_FRB_wml_2024}, the progenitors proposed may emit HE and VHE gamma-rays with mechanisms similar to known PWNe or SNRs.
These models suggest that magnetars can be interesting targets for IACTs such as LST$-$1 and the upcoming arrays of the CTAO \citep{CTAO_galactictransients_2024}.

\subsection{Bursting emission}
\label{sec:discussion:sub:transient}
In Sect.~\ref{sec:TransientEmission:sub:ToA} and Table~\ref{table:BurstUpperLimits} we obtained the VHE ULs to the bursting emission of SGR~1935$+$2154 on a $\SI{0.1}{\second}$ timescale, centred on the ToAs of simultaneous bursts.
Burst fluences and spectral analyses are available in the literature only for the four bursts detected by \textit{Fermi}-GBM and \textit{NICER}, and we can compare their X-ray flux with the simultaneous VHE UL.

\begin{figure}[ht!]
   \centering
   \resizebox{\hsize}{!}{\includegraphics{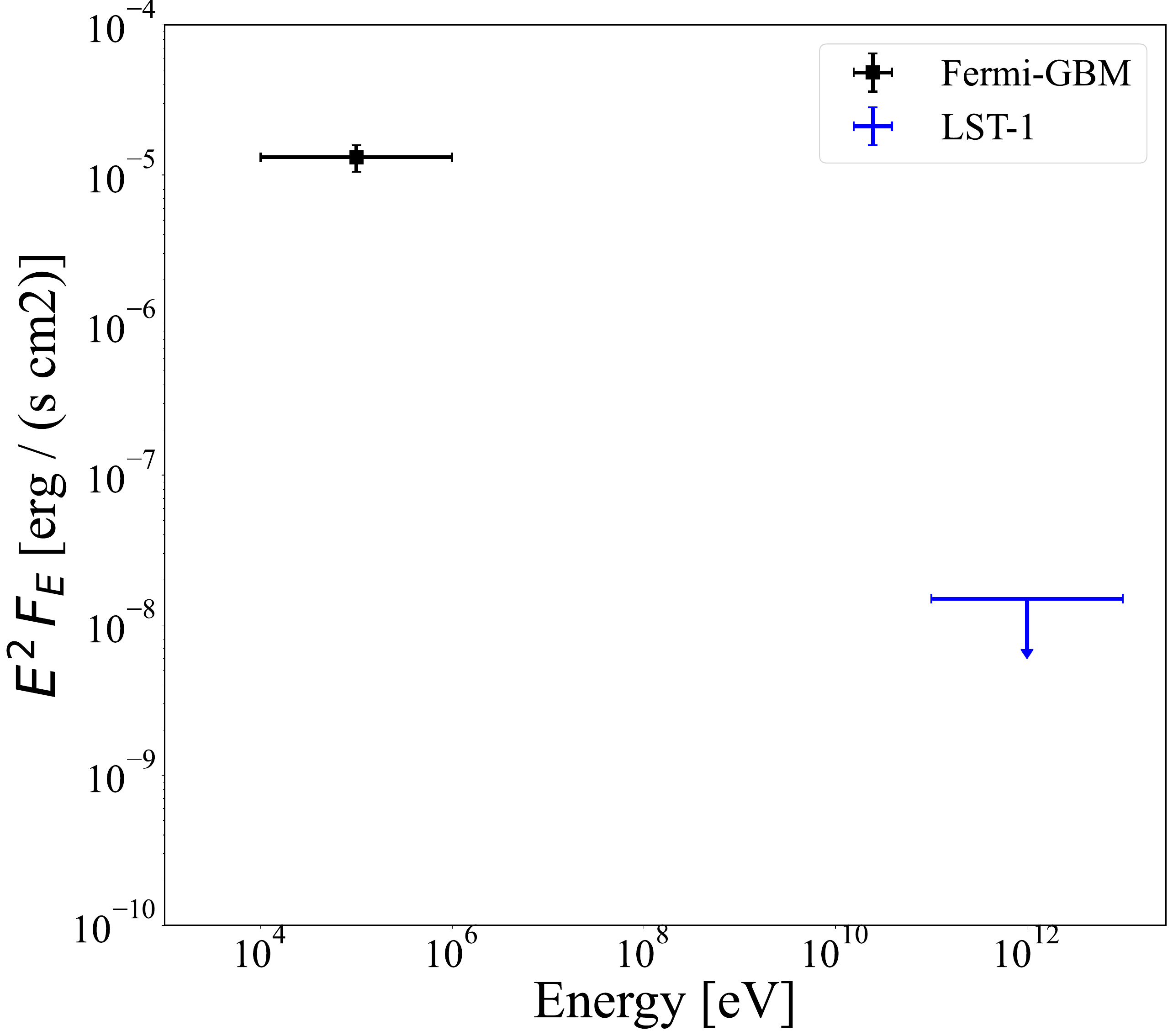}}
   \caption{Soft and VHE gamma-ray SED of the bursting emission of SGR~1935$+$2154 during the burst on July 7 2021 00:33:31.670, measured on a $\sim \SI{0.1}{\second}$ timescale.
   }
 \label{fig:Burst1_SED_E2DNDE}
\end{figure}

The brightest of these bursts is $\#1$, detected by \textit{Fermi}-GBM with $T_0$=2021-07-07 00:33:31.67 \citep{atel_14764_2021, gcn_30407_2021}.
It has a single peak with $T_{90}\approx\SI{0.1}{\second}$ and is best fit by a power-law function with index $\num{0.54(07)}$ and an exponential HE cut-off at $E_\text{peak}=\SI{36.1(3)}{\kilo\electronvolt}$, with an average photon flux of $\SI{289(3)}{\second^{-1}\centi\meter^{-2}}$, average energy flux of $\SI{1.3(2)E-5}{\erg\,\second^{-1}\centi\meter^{-2}}$, and fluence of $\SI{2.52(03)E-6}{\erg\,\centi\meter^{-2}}$ ($\qtyrange{10}{1000}{\kilo\electronvolt}$, $T_0-\SI{0.064}{\second}$ to $T_0+\SI{0.128}{\second}$).
The $\num{0.1}-\num{10}\,\si{\tera\electronvolt}$ flux UL over $\SI{0.1}{\second}$ is $\SI{1.1e-8}{\erg\per\second\per\centi\meter\squared}$ (see Table~\ref{table:BurstUpperLimits}).
We show in Fig.~\ref{fig:Burst1_SED_E2DNDE} the SED of burst $\#1$ using the \textit{Fermi}-GBM flux point and the VHE UL.
It is the most robust SED of a magnetar burst with a sub-second timescale and simultaneous X-ray and VHE data, and it constrains the ratio between the VHE and X-ray flux to a factor of $\lesssim \num{e-3}$.

Most magnetar models predict bursting emission in the X-ray and soft gamma-ray bands, either when magnetic stresses build up sufficiently to crack a patch of the NS crust, ejecting hot plasma into the magnetosphere \citep[e.g.][]{thompson_outbursts_1995}, or after magnetic reconnection events in the magnetosphere \citep{lyutikov_radiomagnetars_2002}.
Such models do not predict gamma-ray emission in the $\si{\giga\electronvolt}-\si{\tera\electronvolt}$ energy ranges, but more complex magnetar models predict mechanism to emit FRBs accompanied by HE $\si{\giga\electronvolt}-\si{\tera\electronvolt}$ bursts. 
For instance, \citet{lyubarsky_frbmaser_2014} predicts that FRBs and strong millisecond bursts in the tera-electronvolt range can originate from the synchrotron maser emission caused by magnetised shocks that occur when highly magnetised plasma bursting from a magnetar reaches the medium surrounding the nebula.

\citet{murase_burst_2016, murase_erratum_2017} suggested a scenario for the production of FRBs in which the spin-down power of a magnetar creates a relativistic wind bubble embedded in baryonic ejecta surrounding the NS, a scenario similar to PWNe.
An impulsive energy injection of relativistic flow, which may originate from magnetic dissipation in the magnetosphere, may shock the nebula, causing the emission of a broadband flare observable from the radio band to the HE gamma-ray band.
The fluence, $\varphi$, predicted for a HE gamma-ray flash (HEGF) in the fast cooling scenario (electrons and positrons cool within the dynamical time) at tera-electronvolt energies is
\begin{equation}\label{eq:MuraseFluence}
    \varphi \sim \SI{8e-8}{\erg \per\centi\meter\squared} \left( \frac{E_\text{outflow}}{\SI{e48}{\erg}}\right) \left( \frac{d}{\SI{e5}{\kilo\parsec}}\right)^{-2},
\end{equation}
where $E_\text{outflow}$ is the total energy of the burst outflow and $d$ is the distance of the magnetar nebula.
A $\si{\tera\electronvolt}$ fluence UL of $\SI{1.1e-9}{\erg\per\centi\meter\squared}$ such as the one we obtained for the SGR~1935$+$2154 burst $\#1$ would constrain the total energy of the burst outflow to $E_\text{outflow} \lesssim \SI{2.7e37}{\erg}$ assuming distance $d=\SI{4.4}{\kilo\parsec}$ \citep{mereghetti_integral_2020}.
We stress that the model cannot be directly applied to burst $\#1$ as it is a typical magnetar burst with no radio counterpart, and a softer index and peak energy with respect to the burst on April 2020 associated with radio bursts \citep{mereghetti_integral_2020}.
Furthermore, SGR~1935$+$2154 lacks the nebular structure that is required for a FRB-HEGF event \citep{murase_burst_2016}.
Nevertheless, other sources of FRBs may possess the structures required for HEGFs, and these theoretical models motivate the search for counterparts of FRBs and magnetar bursts and flares at VHE with IACTs \citep{carosilopezoramas_transientreview_2024}.

The detection of a candidate MGF in the nearby ($\SI{3.5}{\mega\parsec}$) Sculptor galaxy at $\si{\giga\electronvolt}$ energies \citep{ajello_giant_flare_2021} also motivates the search for magnetar flares at HE and VHE.
It was described by a power-law spectrum with index $\num{-1.7(3)}$ and integral flux above $\SI{0.1}{\giga\electronvolt}$ $\displaystyle \Phi \left(>\SI{0.1}{\giga\electronvolt} \right)= (4.1\pm2.2)\,\SI{e-6}{\second^{-1}\centi\meter^{-2}}$ \citep{ajello_giant_flare_2021}.
Assuming that this power-law spectrum extends to tera-electronvolt energies, the estimated average flux above $\SI{0.1}{\tera\electronvolt}$ at the distance of SGR~1935$+$2154 would be approximately $\displaystyle \Phi \left(>\SI{0.1}{\tera\electronvolt} \right) \simeq \SI{2.0e-3}{\second^{-1}\centi\meter^{-2}}$.
The LST$-$1 sensitivity for a $\SI{0.1}{\second}$ burst is $\simeq \SI{3e-8}{\second^{-1}\centi\meter^{-2}}$, suggesting that a short flare with the average flux of the MGF could be detected up to a distance of a few $\si{\mega\parsec}$.
However, the sensitivity can vary significantly depending on the source position (e.g. within or outside the Galactic plane), observing conditions, and flare duration.
Furthermore, this estimate assumes that the spectrum measured by \citet{ajello_giant_flare_2021} extends to tera-electronvolt energies, which remains to be confirmed.

\section{Conclusions}
\label{sec:conclusion}
We analysed over $\SI{25}{\hour}$ of high-quality observations of magnetar SGR~1935$+$2154 with LST$-$1 in order to search for a VHE counterpart to its persistent or bursting emission. We did not detect the persistent emission of the source in the $\qtyrange{0.1}{10}{\tera\electronvolt}$ energy range, in agreement with previous observations \citep{hess_sgr1935_2021}, and as is implied by the $\si{\mega\electronvolt}$ ULs, if there is only a single spectral component extending from X-rays to VHE.
Our tera-electronvolt ULs are about the same order of magnitude as the X-ray emission of SGR~1935$+$2154.

We conducted our observations during periods of known bursting activity of the source, and simultaneously to nine short magnetar bursts detected by instruments sensitive to X-rays on board several space satellites.
We searched for a transient VHE emission of SGR~1935$+$2154 that coincided with the HE bursts on a timescale of $\SI{0.1}{\second}$, in a low-photon-statistics regime and under the hypothesis of a Poisson background, optimising the event selection cuts for the analysis.
For the first time, we provided the VHE flux ULs to the emission of a short magnetar burst ($\sim \SI{1.2e-8}{\erg\per\second\per\centi\meter\squared}$) simultaneous to its soft gamma-ray flux.
For the brightest burst in our sample, the ratio between the VHE and X-ray flux is $\lesssim \num{e-3}$.
We also searched for possible $\SI{0.1}{\second}$ bursts in our dataset and we did not observe any significant signal. 

Although no persistent or bursting emission has yet been detected at tera-electronvolt energies from SGR~1935$+$2154 and other magnetars, the detection of $\si{\mega\electronvolt}-\si{\giga\electronvolt}$ emission and theoretical predictions of a VHE component in flares make magnetars and FRBs interesting candidate sources for IACTs and CTAO in particular.
A flare with the same power-law spectrum as the candidate MGF detected in the Sculptor galaxy \citep{ajello_giant_flare_2021} could have been within the detection capabilities of LST$-$1, assuming that the power law extends up to tera-electronvolt energies.
A next-generation facility such as the CTAO will be particularly suited for fast transient astronomy thanks to its improved VHE sensitivity\footnote{\url{https://www.ctao.org/for-scientists/performance/}} (approximately an order of magnitude greater than that of LST$-$1).
This enhanced sensitivity, primarily attributed to the Large-Sized Telescopes (LSTs) in the low-energy range, will not only improve detection capabilities but also lower the energy threshold, enabling the CTAO to observe gamma-ray fluxes comparable to those expected from magnetar bursts and flares \citep{CTAO_galactictransients_2024}.
The burst analysis method described in Sect.~\ref{sec:TransientEmission} is an alternative to the standard methods of Sect.~\ref{sec:PersistentEmission} and is particularly useful for the analysis of fast transients such as FRBs, magnetar bursts and flares, and short gamma-ray bursts that may belong to a small-photon-statistics regime.
Another key feature of CTAO in the context of transient astronomy will be its capability to swiftly react to external alerts from complementary astrophysical facilities by triggering target-of-opportunity observations.
The CTAO will be able to swiftly re-point its telescopes and perform real-time analyses using its Science Alert Generation system \citep{di_piano_detection_2021, bulgarelli_SAGproceedingsSPIE_2024}, which is part of the Array Control and Data Acquisition of CTAO \citep{oya_acadaspie_2024}.
A real-time analysis system has already been deployed for LST$-$1 to monitor nightly observations \citep{caroff_real_2023}.

\bibliographystyle{aa}
\bibliography{References}

\begin{appendix}
\onecolumn
\section{Solution of the upper limit equation}
\label{sec:appendix_integrals}
For completeness, we provide the intermediate steps needed to obtain the explicit form of the posterior distribution $\Po_\text{sig}(s|n)$ in eq.~\eqref{eq:Casadei_Posterior_Dirac}, starting from its definition in eq.~\eqref{eq:Casadei_Posterior}, and the UL equation eq.~\eqref{eq:bayesianupperlimitpoisson} in Sect.~\ref{sec:TransientEmission:sub:ToA}.
The derivations rely on the definition of the complete Gamma function
\begin{equation}\label{eq:GammaFunction}
    \Gamma (y) = \int_0^\infty \, t^{y-1} e^{-t} \, \d t,
\end{equation}
the lower incomplete Gamma function
\begin{equation}\label{eq:GammaIncompleteLower}
    \gamma (y, x) = \int_0^x \, t^{y-1} e^{-t} \, \d t,
\end{equation}
and their property
\begin{equation}
    \label{eq:Gamma_to_Chi2CDF}
    \gamma(y,x) = \Gamma(y) \cdot F_{\chi^2} (2x, 2y),
\end{equation}
where $F_{\chi^2}$ denotes the cumulative distribution function of a $\chi^2$ distribution with $2y$ degrees of freedom \citep[see e.g.][]{feigelson_babu_statistics_2012}.

\subsection{Posterior Distribution}
\label{sec:appendix_integrals:sub:posterior}
The explicit form of the marginalised likelihood follows from eq.~\eqref{eq:Casadei_Likelihood_marginalized} when the background prior is taken to be a Dirac delta, eq.~\eqref{eq:BackgroundPriorDiracDelta}:
\begin{equation}
\label{eq:Casadei_Likelihood_marginalized_explicit}
    \Li_\text{sig}(n|s)=\Poi(n;s+b_0) = \frac{(s+b_0)^n}{n!} \, e^{-(s+b_0)}.
\end{equation}
We use eq.~\eqref{eq:Casadei_Likelihood_marginalized_explicit} and eq.~\eqref{eq:ON_Prior_Dirac} to evaluate the normalisation in the denominator of eq.~\eqref{eq:Casadei_Posterior}:
\begin{equation}
\label{eq:integral_evidence}
    \begin{split}
        & \int_0^\infty \, \pi_\text{sig} (s)\, \Li_\text{sig}(n|s) \, \d s
        = \int_0^\infty \, \sqrt{\frac{b_0}{s+b_0}}  \frac{(s+b_0)^n}{n!}  e^{-(s+b_0)} \, \d s
        = \frac{b_0^\frac{1}{2}}{n!} \int_0^\infty \, (s+b_0)^{n-\frac{1}{2}}  e^{-(s+b_0)} \, \d s
        \underset{\boxed{t=s+b_0}}{=}
        \frac{b_0^\frac{1}{2}}{n!} \int_{b_0}^\infty \, t^{n-\frac{1}{2}}  e^{-t} \, \d t = \\
        &= \frac{b_0^\frac{1}{2}}{n!} \left( \int_0^\infty \, t^{n-\frac{1}{2}}  e^{-t} \, \d t - \int_0^{b_0} \, t^{n-\frac{1}{2}}  e^{-t} \, \d t \right)
        = \frac{b_0^\frac{1}{2}}{n!} \left[ \Gamma \left(n+\frac{1}{2}\right) - \gamma \left(n+\frac{1}{2}, b_0\right) \right] \underset{\boxed{ \text{Eq.~\eqref{eq:Gamma_to_Chi2CDF} }}}{=}
        \frac{b_0^\frac{1}{2}}{n!} \left[ \Gamma\left(n+\frac{1}{2}\right) - \Gamma\left(n+\frac{1}{2}\right) \cdot F_{\chi^2} \left( 2b_0; 2n+1 \right) \right]\\
        &= \frac{b_0^\frac{1}{2}}{n!} \cdot \Gamma\left(n+\frac{1}{2}\right) \cdot \left[ 1 - F_{\chi^2} \left( 2b_0; 2n+1 \right) \right].
    \end{split}
\end{equation}
Substituting this result into eq.~\eqref{eq:Casadei_Posterior} and simplifying shows that the prefactor $\tfrac{b_0^{1/2}}{n!}$ cancels, leading directly to the posterior in eq.~\eqref{eq:Casadei_Posterior_Dirac}.

\subsection{Upper limit}
\label{sec:appendix_integrals:sub:upperlimit}
The derivation of the Bayesian UL in eq.~\eqref{eq:bayesianupperlimitpoisson} proceeds in a similar way. Starting from its definition, eq.~\eqref{eq:BayesianUpperLimit}, we have
\begin{equation}
\label{eq:ONOFF_UpperLimit_Equation}
    \begin{split}
        1-\alpha_\text{CL}
        &= \int_0^{s_\text{UL}} \, \Po_\text{sig}(s|n) \, \d s
        = \frac{ \int_0^{s_\text{UL}} \, (s+b_0)^{n-\frac{1}{2}} \, e^{-(s+b_0)} \, \d s}{\Gamma \left(n+\frac{1}{2}\right) \cdot \left[ 1 - F_{\chi^2} \left( 2 b_0; 2n+1 \right) \right]}
        = \frac{ \int_{b_0}^{s_\text{UL}+b_0} \, t^{n-\frac{1}{2}} e^{-t} \, \d t}{\Gamma \left(n+\frac{1}{2}\right) \cdot \left[ 1 - F_{\chi^2} \left( 2 b_0; 2n+1 \right) \right]}=\\
        &= \frac{ \int_{0}^{s_\text{UL}+b_0} \, t^{n-\frac{1}{2}} e^{-t} \, \d t - \int_{0}^{b_0} \, t^{n-\frac{1}{2}} e^{-t} \, \d t}{\Gamma \left(n+\frac{1}{2}\right) \cdot \left[ 1 - F_{\chi^2} \left( 2 b_0; 2n+1 \right) \right]}
        = \frac{ \gamma \left(n+\frac{1}{2}, s_\text{UL}+b_0\right) - \gamma \left(n+\frac{1}{2}, b_0\right)}{\Gamma \left(n+\frac{1}{2}\right) \cdot \left[ 1 - F_{\chi^2} \left( 2 b_0; 2n+1 \right) \right]}
        =\frac{ F_{\chi^2} \left( 2 \left(s_\text{UL}+b_0\right); 2n+1 \right) - F_{\chi^2} \left( 2 b_0; 2n+1 \right)}{1 - F_{\chi^2} \left( 2 b_0; 2n+1 \right)}.
    \end{split}
\end{equation}
Rearranging, we obtain the equivalent condition:
\begin{align}
    F_{\chi^2} \left( 2\left(s_\text{UL}+b_0\right); 2n+1 \right) &= \left( 1- \alpha_\text{CL} \right) \cdot \left[ 1 - F_{\chi^2} \left( 2 b_0; 2n+1 \right) \right] + F_{\chi^2} \left( 2 b_0; 2n+1 \right),  \\
    F_{\chi^2} \left( 2\left(s_\text{UL}+b_0\right); 2n+1 \right) &= 1 -\alpha_\text{CL} \cdot \left[ 1 - F_{\chi^2} \left( 2 b_0; 2n+1 \right) \right],\\
    2\left(s_\text{UL}+b_0\right) &= F_{\chi^2}^{-1}\left(1 -\alpha_\text{CL} \cdot \left[ 1 - F_{\chi^2} \left( 2 b_0; 2n+1 \right) \right]; 2n+1\right).
\end{align}
From this result we can determine $s_\text{UL}$ directly, leading to eq.~\eqref{eq:bayesianupperlimitpoisson}.
\end{appendix}

\section*{Acknowledgments}
\label{sec:acknowledgements}
\FloatBarrier
\begin{acknowledgements}
We gratefully acknowledge financial support from the following agencies and organisations:
Conselho Nacional de Desenvolvimento Cient\'{\i}fico e Tecnol\'{o}gico (CNPq) Grant 309053/2022-6 and Funda\c{c}\~{a}o de Amparo \`{a} Pesquisa do Estado do Rio de Janeiro (FAPERJ) Grants E-26/200.532/2023 and E-26/211.342/2021, Funda\c{c}\~{a}o de Amparo \`{a} Pesquisa do Estado de S\~{a}o Paulo (FAPESP), Funda\c{c}\~{a}o de Apoio \`{a} Ci\^encia, Tecnologia e Inova\c{c}\~{a}o do Paran\'a - Funda\c{c}\~{a}o Arauc\'aria, Ministry of Science, Technology, Innovations and Communications (MCTIC), Brasil;
Ministry of Education and Science, National RI Roadmap Project DO1-153/28.08.2018, Bulgaria;
Croatian Science Foundation (HrZZ) Project IP-2022-10-4595, Rudjer Boskovic Institute, University of Osijek, University of Rijeka, University of Split, Faculty of Electrical Engineering, Mechanical Engineering and Naval Architecture, University of Zagreb, Faculty of Electrical Engineering and Computing, Croatia;
Ministry of Education, Youth and Sports, MEYS  LM2023047, EU/MEYS CZ.02.1.01/0.0/0.0/16\_013/0001403, CZ.02.1.01/0.0/0.0/18\_046/0016007, CZ.02.1.01/0.0/0.0/16\_019/0000754, CZ.02.01.01/00/22\_008/0004632 and CZ.02.01.01/00/23\_015/0008197 Czech Republic;
CNRS-IN2P3, the French Programme d’investissements d’avenir and the Enigmass Labex, 
This work has been done thanks to the facilities offered by the Univ. Savoie Mont Blanc - CNRS/IN2P3 MUST computing center, France;
Max Planck Society, German Bundesministerium f{\"u}r Forschung, Technologie und Raumfahrt (Verbundforschung / ErUM), the Deutsche Forschungsgemeinschaft (SFB 1491) and the Lamarr-Institute for Machine Learning and Artificial Intelligence, Germany;
Istituto Nazionale di Astrofisica (INAF), Istituto Nazionale di Fisica Nucleare (INFN), Italian Ministry for University and Research (MUR), and the financial support from the European Union -- Next Generation EU under the project IR0000012 - CTA+ (CUP C53C22000430006), announcement N.3264 on 28/12/2021: ``Rafforzamento e creazione di IR nell’ambito del Piano Nazionale di Ripresa e Resilienza (PNRR)'';
ICRR, University of Tokyo, JSPS, MEXT, Japan;
JST SPRING - JPMJSP2108;
Narodowe Centrum Nauki, grant number 2023/50/A/ST9/00254, Poland;
The Spanish groups acknowledge the Spanish Ministry of Science and Innovation and the Spanish Research State Agency (AEI) through the government budget lines
PGE2022/28.06.000X.711.04,
28.06.000X.411.01 and 28.06.000X.711.04 of PGE 2023, 2024 and 2025,
and grants PID2019-104114RB-C31,  PID2019-107847RB-C44, PID2019-105510GB-C31, PID2019-104114RB-C33, PID2019-107847RB-C43, PID2019-107847RB-C42, PID2019-107988GB-C22, PID2021-124581OB-I00, PID2021-125331NB-I00, PID2022-136828NB-C41, PID2022-137810NB-C22, PID2022-138172NB-C41, PID2022-138172NB-C42, PID2022-138172NB-C43, PID2022-139117NB-C41, PID2022-139117NB-C42, PID2022-139117NB-C43, PID2022-139117NB-C44, PID2022-136828NB-C42, PID2024-155316NB-I00, PDC2023-145839-I00 funded by the Spanish MCIN/AEI/10.13039/501100011033 and by ERDF/EU and NextGenerationEU PRTR; CSIC PIE 202350E189; the "Centro de Excelencia Severo Ochoa" program through grants no. CEX2020-001007-S, CEX2021-001131-S, CEX2024-001442-S; the "Unidad de Excelencia Mar\'ia de Maeztu" program through grants no. CEX2019-000918-M, CEX2020-001058-M; the "Ram\'on y Cajal" program through grants RYC2021-032991-I  funded by MICIN/AEI/10.13039/501100011033 and the European Union “NextGenerationEU”/PRTR and RYC2020-028639-I; the "Juan de la Cierva-Incorporaci\'on" program through grant no. IJC2019-040315-I and "Juan de la Cierva-formaci\'on"' through grant JDC2022-049705-I; the “Viera y Clavijo” postdoctoral program of Universidad de La Laguna, funded by the Agencia Canaria de Investigaci\'on, Innovaci\'on y Sociedad de la Informaci\'on. They also acknowledge the "Atracci\'on de Talento" program of Comunidad de Madrid through grant no. 2019-T2/TIC-12900; “MAD4SPACE: Desarrollo de tecnolog\'ias habilitadoras para estudios del espacio en la Comunidad de Madrid" (TEC-2024/TEC-182) project, Doctorado Industrial (IND2024/TIC34250) and Ayudas para la contrataci\'on de personal investigador predoctoral en formación (PIPF-2023/TEC-29694) funded by Comunidad de Madrid; the La Caixa Banking Foundation, grant no. LCF/BQ/PI21/11830030; Junta de Andaluc\'ia under Plan Complementario de I+D+I (Ref. AST22\_0001) and Plan Andaluz de Investigaci\'on, Desarrollo e Innovaci\'on as research group FQM-322; Project ref. AST22\_00001\_9 with funding from NextGenerationEU funds; the “Ministerio de Ciencia, Innovaci\'on y Universidades”  and its “Plan de Recuperaci\'on, Transformaci\'on y Resiliencia”; “Consejer\'ia de Universidad, Investigaci\'on e Innovaci\'on” of the regional government of Andaluc\'ia and “Consejo Superior de Investigaciones Cient\'ificas”, Grant CNS2023-144504 funded by MICIU/AEI/10.13039/501100011033 and by the European Union NextGenerationEU/PRTR,  the European Union's Recovery and Resilience Facility-Next Generation, in the framework of the General Invitation of the Spanish Government's public business entity Red.es to participate in talent attraction and retention programmes within Investment 4 of Component 19 of the Recovery, Transformation and Resilience Plan; Junta de Andaluc\'{\i}a under Plan Complementario de I+D+I (Ref. AST22\_00001), Plan Andaluz de Investigaci\'on, Desarrollo e Innovación (Ref. FQM-322). ``Programa Operativo de Crecimiento Inteligente" FEDER 2014-2020 (Ref.~ESFRI-2017-IAC-12), Ministerio de Ciencia e Innovaci\'on, 15\% co-financed by Consejer\'ia de Econom\'ia, Industria, Comercio y Conocimiento del Gobierno de Canarias; the "CERCA" program and the grants 2021SGR00426 and 2021SGR00679, all funded by the Generalitat de Catalunya; and the European Union's NextGenerationEU (PRTR-C17.I1). This work is funded/Co-funded by the European Union (ERC, MicroStars, 101076533). This research used the computing and storage resources provided by the Port d'Informaci\'o Cient\'ifica (PIC) data center.
State Secretariat for Education, Research and Innovation (SERI) and Swiss National Science Foundation (SNSF), Switzerland;
The research leading to these results has received funding from the European Union's Seventh Framework Programme (FP7/2007-2013) under grant agreements No~262053 and No~317446;
This project is receiving funding from the European Union's Horizon 2020 research and innovation programs under agreement No~676134;
ESCAPE - The European Science Cluster of Astronomy \& Particle Physics ESFRI Research Infrastructures has received funding from the European Union’s Horizon 2020 research and innovation programme under Grant Agreement no. 824064.

    \\
    \textit{Software used.}
    lstchain v0.9.13 \citep{lstchain_proceeding_2021, lstchain-Zenodo_2023}, 
    Gammapy v1.3 \citep{gammapy_paper_2023, gammapy_software1.3_2025}, 
    lstMCpipe v0.10.0 \citep{lstMCpipe_article_2022, lstMCpipe-Zenodo_2023}, 
    LSTOSA v0.9.2 \citep{ruiz_lstosa_2022, morcuende_lstosa_2022}.
    
    \\
    \textit{Author Contributions.}
    G. Panebianco: project coordination, LST$-$1 data analysis, statistical methods, burst analysis pipeline development, results discussion and interpretation.
    A. L\'opez-Oramas: PI of the proposal, campaign and strategy coordinator.
    S. Mereghetti: results discussion and interpretation, paper drafting and editing.
    A. Simongini: LST$-$1 data analysis.
    A. Bulgarelli: project supervision, paper drafting and editing.
    P. Bordas: project coordination.
    A. Carosi: LST$-$1 data analysis supervision.
    A. Di Piano: statistical methods, paper drafting and editing.
    T. Hassan: burst analysis methods.
    I. Jiménez Martínez, paper drafting and editing.
    R. L\'opez-Coto: LST$-$1 persistent emission data analysis.
    N. Parmiggiani: statistical methods, paper drafting and editing.
    C. Vignali: project supervision, paper drafting and editing.
    R. Zanin: project coordination.
    
    All authors above have participated in the paper discussion and edition.
    The rest of the authors have contributed in one or several of the following ways: design, construction, maintenance and operation of the instrument(s) used to acquire the data; preparation and/or evaluation of the observation proposals; data acquisition, processing, calibration and/or reduction; production of analysis tools and/or related MC simulations; discussion and approval of the contents of the draft.
    We kindly thank the anonymous referee and the CTAO internal reviewers for their careful reading and constructive comments, which helped improve the quality of this work.
\end{acknowledgements}

\end{document}